\documentclass[aps,prb,twocolumn,showpacs,preprintnumbers]{revtex4-1}

\usepackage[T1]{fontenc}
\usepackage[latin9]{inputenc}
\usepackage{amsmath}
\usepackage{graphicx}
\usepackage{amssymb}

\usepackage{CJK}

\makeatother

\begin{document}
\title{Circular Optical Nanoantennas - An Analytical Theory}

\author{Robert~Filter}
\author{Jing~Qi~\begin{CJK}{UTF8}{gbsn}(戚婧)\end{CJK}}
\author{Carsten~Rockstuhl}
\author{Falk~Lederer}
\affiliation{Institute of Condensed Matter Theory and Solid State
Optics, Abbe Center of Photonics, Friedrich-Schiller-Universit\"{a}t Jena, Max-Wien-Platz 1, D-07743 Jena, Germany}

\begin{abstract}
An analytical approach is provided for describing the resonance properties of optical nanoantennas made
of a stack of homogeneous discs, i.e. circular patch nanoantennas. It consists in analytically calculating the
phase accumulation of surface plasmon polaritons across the resonator and an additional contribution from the
complex reflection coefficient at the antenna termination. This makes the theory self-contained with no need for
fitting parameters. The very antenna resonances are then explained by a simple Fabry-Perot resonator model.
Predictions are compared to rigorous simulations and show excellent agreement. Using this analytical model,
circular antennas can be tuned by varying the composition of the stack.
\end{abstract}

%\doi{Inserted by Editor}

\maketitle

\section{Introduction}

Ever since Hertz described in 1887 the emission of electromagnetic radiation from dipole antennas,
a pertinent question has been to fabricate them such that they may interact with light at optical frequencies,
a spectral domain of paramount importance for many applications.
To achieve this goal antennas have to be downscaled in their critical dimensions to a few hundreds of nanometers
to match the wavelength of the visible. More than a century later, this spectral domain has been reached owing to
recent advances in nanofabrication and characterization techniques\cite{Muhlschlegel2005}. However,
despite of this progress, the theoretical understanding of optical nanoantennas is lagging behind the available
technology. The complexity of this issue is also one of the reasons why the topic attracts much research interest\cite{Novotny2008}.
Whereas metallic antennas at radio frequencies, where the metal may be considered as perfect conductor, can be
conveniently treated analytically using Babinet's principle, this ceases to hold in the visible\cite{Zentgraf2007,Novotny2011,Giannini2011}.
There, the dielectric functions of metals have to be properly accounted for and may be phenomenologically described
by the free-electron (plasma) model.

The arising plasma oscillations may couple to the electromagnetic radiation at a metallic interface to form new
confined quasi-particles \cite{Sernelius2001,Maier2007}. These quasi-particles are termed
surface plasmon polaritons (SPPs) and govern the resonant behavior of optical nanoantennas\cite{Alu2008}.
Such optical nanoantennas enabled already ubiquitous applications
\cite{Rockstuhl2006,Nelayah2007,Anker2008a,Koenderink2009,Giannini2010,Zentgraf2010,Dregely2011,Greffet2011}
and will certainly find many future ones.

Eigenmodes of various nanoantennas have been measured by different means\cite{Vogelgesang2010,Hecht2010,Schnell2010,Yang2011,Hillenbrand2011,Linden2011,Giessen2011,McLeod2011,Giessen2011b}.
However, problematic in further advancing the field is the lack of analytical insight into
the scaling behavior of optical nanoantennas to carefully design them for a desired application\cite{Greffet2005,Esteban2010,Engheta2011,Martin2011,Quidant2011}.
Most design related tasks rely entirely on numerical tools; hence providing only little insight into the underlying physics.
This is in stark contrast to the field of antennas at radio frequencies and it would be highly desirable to bridge this gap.

\begin{figure}
\begin{centering}
\includegraphics[width=80mm]{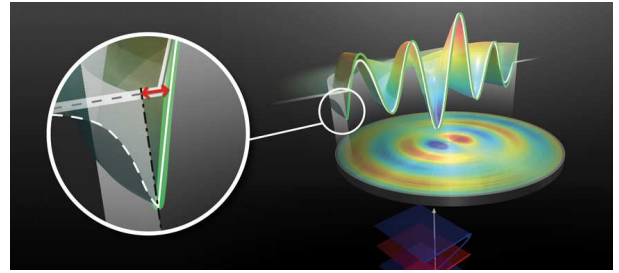}
\par\end{centering}
\caption{\label{fig:ToC}
The excitation of a Bessel-type SPP at a specific circular nanoantenna, a metallic disc.
The reflection at the termination leads to an apparent length change as described by the given theory.}
\end{figure}

First efforts in this direction has been done
characterizing the resonance behavior of the potentially simplest
optical nanoantenna, i.e. the nanowire antenna.
It was shown that their resonances can be explained by a Fabry-Perot
model. Envisaging the nanowire as a plasmonic cavity, a resonance occurs whenever the phase accumulated upon a single round trip
amounts to a multiple of $2\pi$. This phase is determined by the propagation constant of the plasmonic mode supported by the
nanowire and an additional phase upon reflection at the wire termination. By accounting for that additional phase,
the length of the nanoantenna as perceived optically differs from its geometrical length. This phenomenon is sometimes termed the apparent length change of the nanoantenna which turned out to be essential
since it strongly affects the resonance positions\cite{Novotny2007b, Sondergaard2007, Sondergaard2008, Barnard2008, Dorfmuller2009, Gordon2009, Taminiau2011}.
In the past this additional phase contribution has been considered either \textit{ad hoc} without further justification or as
a free parameter in an adapted model. By fitting many resonance positions, taken from many samples with different nanowire lengths and/or incidence angles, to predictions from the model, the phase change upon reflection could be phenomenologically determined. Such tedious procedures were necessary since the analytical calculation of the complex reflection coefficient turned out to be
fairly involved. To date such results are only available for a few simplified geometries\cite{Gordon2006,Gordon2009}.
Understanding and predicting the complex reflection coefficient of surface plasmons at abrupt interfaces is one of the key
issues to be solved for a future advance in nanophotonics.

Our contribution towards this goal is twofold in considering here circular patch nanoantennas, i.e. nanoantennas made from an arbitrary
stack of homogeneous discs.
First, the complex reflection coefficient of Hankel-type surface plasmon polaritons at the circumference of the circular
patch nanoantenna is analytically calculated. This consideration is extremely versatile since many other, but simpler
geometries, can be analyzed while probing for limiting cases of this theory. For example, increasing the radius of a single disc
towards infinity and increasing its thickness yields the reflection coefficient of an SPP propagating along a single metallic
interface at a planar termination.

Moreover, this theory can be used to predict and engineer the resonances
of circular patch nanoantennas made of an, in principle, arbitrary sequence of layers. This is achieved by a Bessel-type standing wave resonator
model using the calculated phase of reflection (see Fig.~\ref{fig:ToC} for a graphical illustration).
Analytical results will be compared to both known results and rigorous simulations to prove the applicability of this theory.

With this work a simple analytical tool is provided that allows for a deeper insight into the scaling behavior of complex
nanoantennas and, moreover, defines a path how to treat such nanooptical systems analytically where only numerical approaches have been ruling in the past.

\section{Reflection of Hankel-type SPPs}

\begin{figure}
\begin{centering}
\includegraphics[width=85mm]{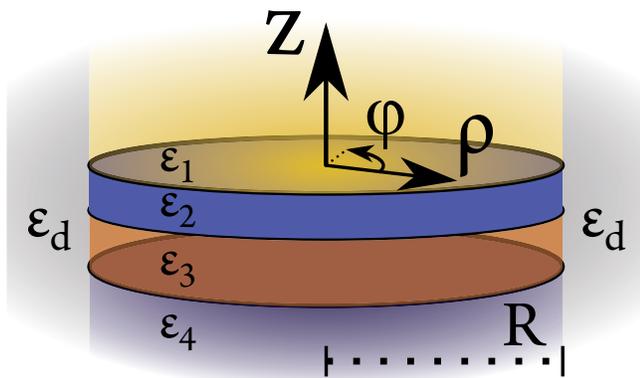}
\par\end{centering}
\caption{\label{fig:axialStack}Schematic of a circular nanoresonator. Its permittivity $\epsilon\left(z\right)$ is axially
symmetric and piecewise homogeneous. The whole structure is radially terminated at $\rho = R$ and embedded in a medium of
permittivity $\epsilon_{d}$.}
\end{figure}

The class of nanoantennas discussed in this work is illustrated
in Fig.~\ref{fig:axialStack}.
Such axially symmetric structures support radially
propagating SPPs which are described by Hankel functions and consequentially termed Hankel-type SPPs\cite{Nerkararyan2010}. In this section, the reflection coefficient of such SPPs at the circumference of axially symmetric nanoantennas will be introduced. Because of the singularity of the Hankel functions at the origin, this ansatz cannot be used to describe the actual modes of circular nanoantennas which will be discussed in the following section.

Assuming a linear response of all involved materials it is sufficient to consider time-harmonic solutions with a $\exp\left(-\mathrm{i}\omega t\right)$ time-dependence.
Furthermore, the plasmonic modes obey an angular variation of $\exp\left(\mathrm{i}m\varphi\right)$ where $m$ denotes the angular mode index. In the following, the time dependency will be dropped and the angular dependency will be denoted by the subscript $m$.

Assuming a transverse-magnetic (TM, $H_{z}=0$) field, it is
sufficient to describe the fields in terms of $E_{z}$ alone.
The plasmonic modes obey the usual boundary conditions at discontinuities
of $\epsilon\left(z\right)$ and the propagation constant in radial direction is subject to a dispersion
relation $k_{\rho}(\omega)\equiv k_{\mathrm{SPP}}(\omega)$.

For the given time-dependency, the Hankel functions $H_{m}^{1/2}$ of the first and second kind correspond to outwards and inwards propagating cylindrical waves. For Hankel-type SPPs, the radially propagating parts of $E_{z}$ are then described by a superposition of the $H_{m}^{1/2}$.

\begin{figure}
\begin{centering}
\includegraphics[width=50mm]{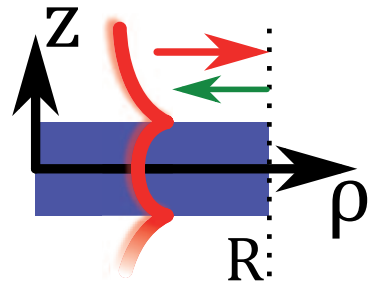}
\par\end{centering}
\caption{\label{fig:ReflectionExplanation}
A Hankel-type SPP\cite{Nerkararyan2010} propagates outwards a circular nanoantenna and gets reflected at its termination into an inwards propagating Hankel-type SPP.}
\end{figure}

Outwards propagating modes are reflected into inwards propagating ones at the circumference of the nanoanantenna. The approach for calculating the corresponding reflection coefficients $r_{m}$ along with the results will be briefly presented in the following. The detailed derivation can be found in App.~\ref{sec:Derivation_of_the_Refl_Coeff}. A schematic of the reflection is given in Fig.~\ref{fig:ReflectionExplanation}.

For the described reflection, the radial part of the surface plasmon field can be described by
\begin{eqnarray}
E_{z}^{m,-}\left(\rho,z\right) & = & \mathcal{A}_{m}\left(k_{\mathrm{SPP}}\rho\right)\cdot a\left(z\right)\ \mathrm{with}\nonumber \\
\mathcal{A}_{m}\left(k_\mathrm{SPP}\rho\right) & = & H_{m}^{1}\left(k_\mathrm{SPP}\rho\right)+r_{m}\cdot H_{m}^{2}\left(k_{\mathrm{SPP}}\rho\right)\label{eq:Ansatz_Ez}\end{eqnarray}
for $\rho\leq R$ implied by the superscript $-$; $+$ will be used later to denote $\rho>R$, the region outside the nanoantenna. The mode profile $a\left(z\right)$ determines the propagation constant $k_{\mathrm{SPP}}$. Furthermore, It is sufficient to calculate $r_m$ for a fixed angular mode $m$ since there will be no reflection into modes with a different rotational symmetry.

It must be noted that this description implicitly neglects the reflection into other eigenfunctions of the resonator and that for
strong absorption this Hankel-type form ceases to be valid; being essentially replaced by Norton waves as recently described\cite{Nikitin2009,Nikitin2010} and further discussed in section \ref{sec:Spectral_Study}.

Outside the resonator the field couples, as in the case of reflection, only to modes of the same angular
symmetry class. A suitable representation is a continuum of outward
propagating Hankel waves \begin{eqnarray*}
E_{z}^{m,+}\left(\mathbf{r}\right) & = & \int_{-\infty}^{\infty}c_{m}\left(k_{z}\right)H_{m}^{1}\left(\sqrt{\epsilon_{d}k_{0}^{2}-k_{z}^{2}}\rho\right)e^{\mathrm{i}k_{z}z}dk_{z}
\end{eqnarray*}
with the corresponding expressions for $H_{\varphi}^{m,+}$.
The amplitudes $c_{m}\left(k_{z}\right)$ are given by obeying the boundary conditions and any scattering of objects outside the nanoantenna e.g. by Bragg mirrors\cite{Benisty1998} is neglected.

Following Gordon\cite{Gordon2006}, the reflection coefficients $r_{m}$ are determined using the
continuity of $H_{\varphi}$ and $\int E_{z}\cdot H_{\varphi}dz$ at $\rho=R$.
Thus, the complex reflection coefficients read as:
\begin{eqnarray}
r_{m} & = & \frac
{2\pi\epsilon_{d}\,k_{\mathrm{SPP}}\sigma\,H_{m}^{1}\!\left(k_{\mathrm{SPP}}R\right)-DH_{m}^{1}\!\left(k_{\mathrm{SPP}}R\right)I_{m}}
{-2\pi\epsilon_{d}\,k_{\mathrm{SPP}}\sigma\,H_{m}^{2}\!\left(k_{\mathrm{SPP}}R\right)+DH_{m}^{2}\!\left(k_{\mathrm{SPP}}R\right)I_{m}}\qquad
\label{eq:reflectionCoeff_most_general}\end{eqnarray}
with the abbreviations
\begin{eqnarray*}
DH_{m}^{1/2}\left(x\right) & \equiv & \partial_{x}H_{m}^{1/2}\left(x\right) \ ,\ \sigma \equiv \int_{-\infty}^{\infty}\epsilon\left(z\right)a\left(z\right)^{2}dz\ ,\\
I_{m} & \equiv & \int_{-\infty}^{\infty}\frac{H_{m}^{1}\left(\sqrt{\epsilon_{d}k_{0}^{2}-k_{z}^{2}}R\right)}{DH_{m}^{1}\left(\sqrt{\epsilon_{d}k_{0}^{2}-k_{z}^{2}}R\right)}\\
 &  & \cdot\sqrt{\epsilon_{d}k_{0}^{2}-k_{z}^{2}}\cdot B^{-}\left(k_{z}\right)\cdot B^{+}\left(k_{z}\right)dk_{z}\ \mathrm{and}\\
B^{\pm}\left(k\right) & \equiv & \int_{-\infty}^{\infty}\epsilon\left(z\right)a\left(z\right)e^{\pm\mathrm{i}kz}dz\ .
\end{eqnarray*}

To compute the phase change upon reflection $\phi_{m}^r$, the
origin of the coordinate system has to be specified. As a result the phase directly computed
from $r_{m}$ is not the phase upon reflection, but an overall accumulated
phase $\phi_{m}^{acc}$ with respect to the origin of the coordinate system.
This overall phase can be decomposed into a reflection $\phi_{m}^{r}$ and a propagation contribution $\phi_{m}^{p}$, thus
$r_{m}  = \left|r_{m}\right|\exp\left(\mathrm{i}\phi_{m}^{acc}\right)$ with $\phi_{m}^{acc}  = \phi_{m}^{r} + \phi_{m}^p$~.
For Hankel-type fields, one finds
\begin{eqnarray*}
\phi_{m}^p & = & \phi_{m}^p\left(0\rightarrow R\right) + \phi_{m}^p\left(R\rightarrow 0\right)\\
& = &
\mathrm{Arg}\left[H_{m}^{1}\left(k_{\mathrm{SPP}}\cdot 0^{+}\right)^{\star}
\cdot   H_{m}^{1}\left(k_{\mathrm{SPP}}\cdot R\right)\right]
+\\&  &
\mathrm{Arg}\left[H_{m}^{2}\left(k_{\mathrm{SPP}}\cdot R\right)^{\star}
\cdot H_{m}^{2}\left(k_{\mathrm{SPP}}\cdot 0^{+}\right) \right]
\end{eqnarray*}
where a further decomposition of the phase accumulation into in- and
outwards propagating waves was used.
Equation~\ref{eq:reflectionCoeff_most_general} is the main analytical result of
this work and constitutes the desired analytical expression for the complex reflection coefficient.

\section{A Simple Resonator Model}

To calculate the complex reflection coefficient it was necessary to use Hankel-type SPPs. Otherwise, in- and outwards propagating waves could not be defined.
Unfortunately, these functions are not finite at the origin and thus not meaningful as standing-wave solutions for $E_z$ inside the resonator.
Thus, to describe the actual field one has to use a plasmonic mode that takes finite values at the origin.
These modes are Bessel-type SPPs described in analogy to Eq.~\ref{eq:Ansatz_Ez} via
\begin{eqnarray}
E_{z}^{m,-}\left(\rho,z\right) & = & J_{m}\left(k_{\mathrm{SPP}}\rho\right)\cdot a\left(z\right)\ .\label{eq:Bessel-Form-Ez}
\end{eqnarray}
The question is how one can combine these two descriptions in the sense of interpreting propagating SPPs as localized ones\cite{Hasan2010}.
As noted earlier, the phase change upon reflection of one-dimensional nanoantennas
causes a deviation of the resonance condition.
The apparent length change was explained by the reflection phase and the same approach will be applied here.
For a reflection without phase accumulation, the resonant radii $R_{n,m}$ of order $n$
would be given by the roots of $E_{z}^{m,-}$ leading to the roots of Bessel functions as for circular microstrip patch antennas \cite{Balanis2005}.
Hence, introducing a nonzero phase upon reflection, one may conjecture the real-valued resonance condition
\begin{eqnarray}
2\cdot k_{\mathrm{SPP}}^{\prime}R_{n,m} + \phi_{m}^{r} & = & 2\cdot x_{n}\left(J_{m}\right)\label{eq:Bessel-resonance-condition}
\end{eqnarray}
where $x_{n}$ denotes the $n$-th root of $J_{m}$, $k_{\mathrm{SPP}}=k_{\mathrm{SPP}}^{\prime}+\mathrm{i}k_{\mathrm{SPP}}^{\prime\prime}$.
The given formulation is a natural extension of the Fabry-Perot resonance condition for one-dimensional nanowire antennas \cite{Dorfmuller2009}. There, the SPP's were characterized by guided plasmonic waves confined by the nanowire and whose propagation along the nanowire is characterized by a propagation constant $k_{\mathrm{SPP}}$. Resonances were found by solving $2k_{\mathrm{SPP}}^{\prime}L_n + 2\phi^r = 2\pi n$. It must be noted that for reflection amplitudes that strongly deviate from unity, a feasible description of the excitation in terms of localized plasmonic modes might not be possible anymore. Thus the employed resonator model loses its predictive strength.

\section{Verification of the Theory}

The latter two equations for the actual form of the fields and for the resonances of circular patch nanoantennas are just educated guesses at this point that have to be verified.

Although different numerical methods could have been used to achieve this\cite{Eremina2011,Urbach2011,Hafner2011}, finite-difference time-domain (FDTD) simulations\cite{Taflove,Oskooi2010a} were carried out for the following subsections. The obtained results were compared to theoretical predictions.

In the first subsection, the assumed Bessel form and scaling of $E_z$ is studied and a comparison to the predictions for resonant radii is performed.
Metallic discs of different thicknesses are used as circular nanoantennas at a fixed frequency. A very good agreement is shown.
The resonator model can further be used to calculate the resonance frequencies of a circular nanoantenna with fixed geometrical parameters. An agreement to simulations is also found in this case and outlined in the second subsection. Furthermore, it is shown how derived quantities can be used to calculate the quality factor ($Q$-factor) of the nanoantenna. Limitations of the theory are discussed.
The section will finish with a study of a stacked structure consisting of two silver discs with a cavity in-between. It will be shown that the theory is also able to accurately describe this situation. This subsection will prove that the theory is versatile and that it can be applied also to circular nanoantennas with a larger complexity.

The convergence to known results in limiting cases is outlined in App.~\ref{sec:Comparison_to_Previous_Results}.

\subsection{\label{sub:Verification_Bessel_and_Scaling}Eigenfunctions and Scaling}

Equations~\ref{eq:Bessel-Form-Ez} and \ref{eq:Bessel-resonance-condition} imply two different things. First of all, the radial dependence $E_z$ is assumed to be described by a Bessel function with a certain scaling caused by the dispersion relation. Second, resonant radii are predicted for which the plasmonic field enhancement is strongest. Both assumptions will be tested in this subsection at a fixed frequency where the permittivity of the metal is just a constant and the observed effects are independent of any nonlinear spectral scaling of $k_{\mathrm{SPP}}(\omega)$.

Discs of several thicknesses $d_i$ and radii between $100\,$nm
and $1250\,$nm with a relative permittivity $\epsilon_{m}\approx-8.8+0.03\mathrm{i}$ were illuminated by a plane wave propagating normally to the disc surface and having a frequency of $\nu=625\,$THz.
The material parameters correspond to those of silver at this frequency\cite{Johnson1972}. The imaginary part has been deliberately lowered to simplify the identification of the resonances. However, for all subsequent subsections the values were directly taken from the literature.
Because of the known wave vector mismatch, the plasmonic modes on planar discs do not couple to those of free space. Hence, the excitation of the SPPs generally takes place at
the disc's edges. Furthermore, a certain mode can only be excited if the outer field has approximately the same spatial symmetry  where the coupling strength is largest. Although the
discs are sufficiently thin such that in general even and odd propagating surface plasmons are supported, the latter ones could not be excited since the disc thicknesses
were much smaller than the wavelength. The required antisymmetric field distribution along the discs thickness does not match to the incident plane wave which shows a phase variation on length scales in the
order of the wavelength. Hence, only even SPPs could be excited with a reasonable efficiency.

\begin{figure}
\begin{centering}
\includegraphics[width=80mm]{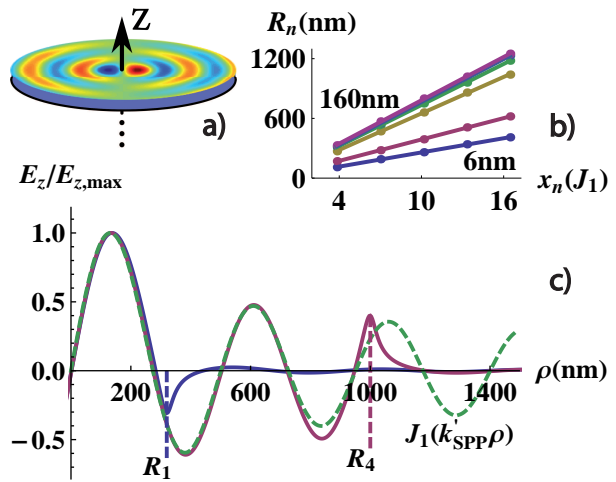}
\par\end{centering}
\caption{\label{fig:Simple-resonator-figs}
Numerical verification of the circular Fabry-Perot resonator model.
a) The electric field component $E_z$ on a metallic disc at resonance shows a qualitative agreement with the assumed Bessel-type form.
b) Resonant radii $R_{n}$ for disc thicknesses of $6$~nm, $10$~nm, $20$~nm, $40$~nm, $80$~nm, and $160$~nm are linearly related to the roots of $J_{1}$.
c) The normalized field $E_z$ on a metallic disc for the first (blue line) and fourth (magenta line) resonant radius.
Inside the discs $\rho<R$ the agreement with the expected form $\propto J_{1}\left(k_{\mathrm{SPP}}\rho\right)$ (green dashed line) is perfect
except a minor deviation at the termination. $k_{\mathrm{SPP}}$ was calculated from the known dispersion relation for a metallic slab\cite{Sernelius2001,Maier2007}.}
\end{figure}

The resonance radii $R_{n}\equiv R_{n,1}$ were determined by identifying
maxima of the electric field strengths below and above the structure
while changing the radius of discs at a constant thickness.
The verification of the employed resonator model requires that a) the excited
fields exhibit the assumed form and b) the measured $R_{n}$ are linearly
related to the roots of $J_{1}$. Looking at Fig.~\ref{fig:Simple-resonator-figs},
it can be recognized that both conditions are met. There, the fields as predicted by Eq.~\ref{eq:Bessel-Form-Ez}
are compared to full wave simulations and an excellent agreement is observed. The fields at nanoantenna resonances
follow exactly such Bessel-type SPPs. Moreover, the resonance radii are linearly related to the roots of $J_{1}$
as assumed in Eq.~\ref{eq:Bessel-resonance-condition}. Consequently, Eqs.~\ref{eq:Bessel-Form-Ez}~and~\ref{eq:Bessel-resonance-condition}
seem to provide an appropriate description of the actual physical situation.

\begin{figure}
\begin{centering}
\includegraphics[width=80mm]{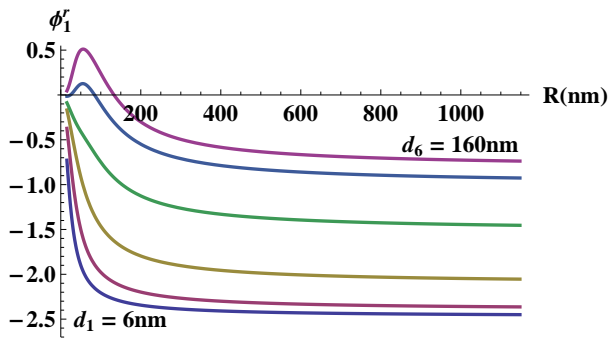}
\par\end{centering}
\caption{\label{fig:Phases_theory}
Phases upon reflection vs. the disc radius calculated from the theoretical model for metallic discs of the same thicknesses as in Fig.~\ref{fig:Simple-resonator-figs}.
The $\phi_{1}^r$ turn out almost always negative. With respect to the existing terminology\cite{Dorfmuller2009},
this corresponds to an apparent length decrease. The effect increases for thinner discs.
}
\end{figure}

The most interesting question might be if the resonant radii $R_n$ can be
explained by an analytically calculated phase of the reflection coefficient
for the excited even modes by using Eq.~\ref{eq:Bessel-resonance-condition} and thus the phases
of the complex reflection coefficient.
The calculated phases $\phi_{1}^r$ can be seen in Fig.~\ref{fig:Phases_theory}.
Phases corresponding to different thicknesses are significantly different. For the given results, it seems to be a monotonic
behavior between thickness $d$ and phase upon reflection $\phi_{1}^r$
- the thicker the disc the larger the reflection phase,
$\phi_{1}^r\left(d_{1}\right)<\phi_{1}^r\left(d_{2}\right)<\dots$
for $d_{i}<d_{i+1}$. Furthermore, all $\phi_{1}^r$ turn out to be
negative, the only exception arises by the two thickest
discs for small radii.
Noteworthy, the relative change of $\phi_{1}^{r}$ from the first
to the fifth resonance radius amounts approximately $10\%$ to $30\%$.
It is increasing with thickness.

\begin{figure}
\begin{centering}
\includegraphics[width=80mm]{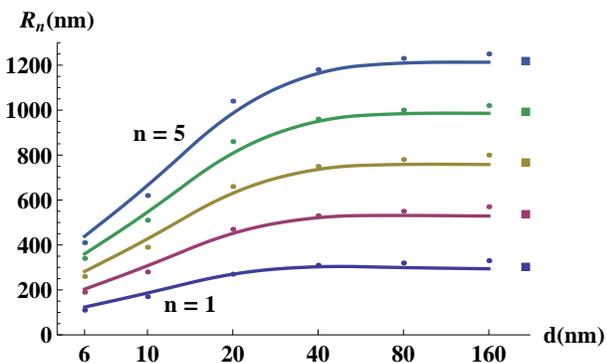}
\par\end{centering}
\caption{\label{fig:Jing_Rn_compare_Sinh_Infinite}
Analytically calculated phases of the reflection coefficient are used to predict the resonance radii $R_{n}$
employing Eq.~\ref{eq:Bessel-resonance-condition} (full lines) for discs of different thicknesses $d$. The five lowest order resonances
are shown. Results from rigorous simulations are shown by dots. Large squares correspond to $R_{n}$ resulting
from calculations assuming a semi-infinite disc; hence the thickness $d$ is infinite.}
\end{figure}

The comparison of numerically and analytically calculated resonant radii is displayed in Fig.~\ref{fig:Jing_Rn_compare_Sinh_Infinite}.
The predictions for even modes coincide very nicely with numerical simulations for thicknesses up to $80\,$nm. This suggests that for a given
thickness the theory predicts a series of resonance radii which are in full agreement with those found using numerical simulations. For an
increasing thickness one can observe that the resonance radii are independent of the thickness. This can be explained while considering the
characteristic penetration depths of the fields into the metallic discs which are in the order of $20\,$nm. Evidently, for thicker discs the
fields confined at the top and the bottom interface are decoupled. Thus the reflection coefficients coincide then with those of semi-infinite
discs. This explains why for thicknesses exceeding $40\,$nm the resonant radii agree very well with the predictions of this approximation
(see App.~\ref{sec:Half-Infinite-cylinder-rm} for an explicit calculation of $r_{m}$ in this case).

\subsection{\label{sec:Spectral_Study}Spectral Dependence of Resonance Frequencies}

In the last subsection the comparison of theoretical predictions to results from simulations was discussed at a
fixed frequency to avoid an inevitable nonlinear scaling due to the frequency dependence of $k_{\mathrm{SPP}}\left(\omega\right)$.
So, it is natural to ask if also the spectral resonant behavior of the circular nanoantennas is in agreement with the given model considering a fixed geometry. Furthermore, using the permittivity of a real metal, one can question what happens if reasonable losses at high frequencies are properly considered. Then, the theory should cease to be valid since mode profiles will differ from the assumed Bessel-type SPP.

\begin{figure}
\begin{centering}
\includegraphics[width=80mm]{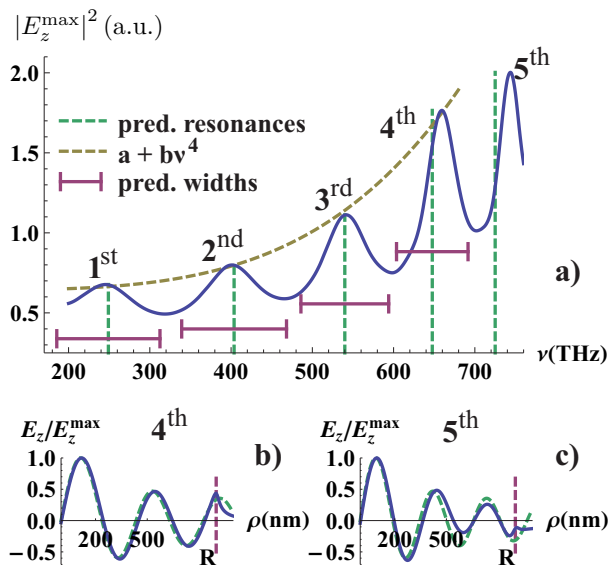}
\par\end{centering}
\caption{\label{fig:FreqScan}
Comparison of theoretical predictions to simulation results for the resonance frequencies of a silver disc with radius $R=900$~nm and thickness $d=20$~nm. a) The maximum value of $\left|E_z\right|^2$ as calculated from rigorous simulations as a function of the illumination frequency $\nu$ (blue line). Resonances (dashed green lines) and line widths (magenta solid line) as predicted by the theory are additionally shown.
For the first four resonances, one can observe a $\nu^4$ scaling (yellow dashed lined). This behavior very likely corresponds to a variation of the incoupling efficiency which is beyond the scope of this work. Nevertheless it is worth mentioning that the scaling does not hold anymore for the fifth resonance and may be explained by the high damping. It furthermore indicates a need to adjust the theoretical description of the system in the respective regime. This is underlined by the actual plasmonic fields:
b) The field distribution at $\varphi=0$ for the fourth order resonance at $\nu\approx660$~THz (blue line) obeys the assumed Bessel-type form (green dashed line).
In c) the situation is shown for the fifth order resonance at $\nu\approx740$~THz. One can observe clear differences to the assumed form which can be attributed to a larger damping of the Bessel-type SPP.}
\end{figure}

To answer these two questions, rigorous simulations were performed for silver discs where the material data was taken form literature\cite{Johnson1972}. The discs were assumed to have a radius of $R=900$~nm and a thickness of $d=20$~nm. They were illuminated by plane waves at different frequencies. Figure~\ref{fig:FreqScan} a) shows the field maxima above the structure as calculated from FDTD\cite{Taflove,Oskooi2010a} simulations as a function of the frequency. In addition the predicted resonance frequencies from the analytical theory are shown. An excellent agreement for nearly all orders can be noticed; although the model is slightly less accurate for higher order resonances. With respect to these results one finds first deviations for the fourth order mode at $\nu\approx660$~THz. There, the permittivity of silver is $\epsilon_m\left(660~\mathrm{THz}\right)\approx-7.3 + 0.2~\mathrm{i}$ leading to a wave vector of $k_{\mathrm{SPP}}\approx\left(1.6\cdot10^7 + 1.2\cdot10^5~\mathrm{i}\right)~\mathrm{m^{-1}}$ for even modes. The propagation length $d_{\mathrm{SPP}} = 1/k^{\prime\prime}_{\mathrm{SPP}} $ is in this case approximately $8.6\mathrm{~\mu m}$ which is roughly five times the diameter of the disc. Thus, this damping is not negligible but with respect to the dimensions of the disc, the resonator model can be applied.

The situation changes at the fifth order resonance at $\nu\approx740$~THz. There, $\epsilon_m\left(740~\mathrm{THz}\right)\approx-4.6 + 2.1~\mathrm{i}$ and $k_\mathrm{{SPP}}\approx\left(2.0\cdot10^7 + 3.5\cdot10^5~\mathrm{i}\right)~\mathrm{m^{-1}}$ leading to a propagation length of $d_{\mathrm{SPP}} \approx 2.8\mathrm{~\mu m}$ which is in the order of the disc diameter. In this highly damped case, the resonance differs from the assumed Bessel-type SPP and the theory is no longer valid\cite{Nikitin2009,Nikitin2010}. This can be also seen comparing the field amplitudes across the surface of the disc as shown in Figs.~\ref{fig:FreqScan} b) and \ref{fig:FreqScan} c) for the fourth and fifth order mode.

\begin{figure}
\begin{centering}
\includegraphics[width=80mm]{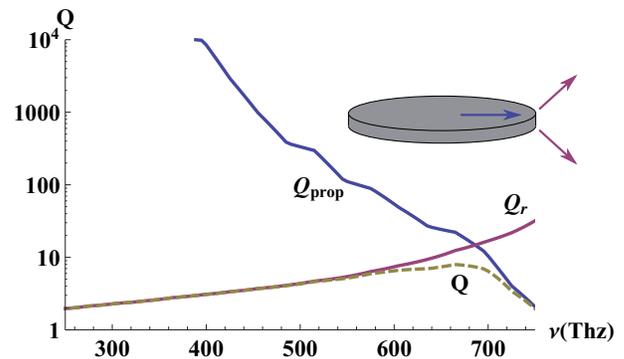}
\par\end{centering}
\caption{\label{fig:QualityFactor}
The quality factor $Q$ as calculated from Eq.~\ref{eq:QualityFactor} for a silver disc with radius $R=900$~nm and thickness $d=40$~nm.
Up to $600$~THz, losses due to propagation are much lower than radiative ones such that $Q$ is entirely dominated by the contribution of $Q_\mathrm{r}$ (magenta line).
The reflection coefficient $r$ is increasing with frequency in this spectral domain. Thus, $Q_\mathrm{r} = \left|1-r\right|^{-2}$ shows the same behavior leading to narrowed peak widths as observed in Fig.~\ref{fig:FreqScan} a).
}
\end{figure}

Looking at the resonance spectrum in Fig.~\ref{fig:FreqScan} a), one can pose the question if the theory can predict the width $\Delta \nu$ of the resonance peaks. Interpreting the Bessel-type SPP as damped oscillation across the nanoantenna which has been done for other kinds of resonant nanostructures\cite{Petschulat2008}, this question directly leads to the quality factor $Q$\cite{Jackson98}. To calculate $Q$, the energy loss of the plasmonic mode per cycle has to be computed. Two direct contributions are given by the theory, propagation losses due to the imaginary part of $k_{\mathrm{SPP}}$ and radiation losses upon reflection.
One finds
\begin{eqnarray}
\frac{1}{Q} & = & \frac{1}{Q_{\text{prop}}} + \frac{1}{Q_\mathrm{r}}\nonumber\\
& = &  \left( 2 k^{\prime\prime}_{\mathrm{SPP}} R \right)^2+ \left|1-r\right|^2 = \frac{\Delta\nu}{\nu}\ .\label{eq:QualityFactor}
\end{eqnarray}
Using Eq.~\ref{eq:QualityFactor}, the quality factor for the Bessel-type SPP was calculated.
The results are displayed in Fig.~\ref{fig:QualityFactor}. Up to the fourth resonance, the quality factor is dominated by the radiative contribution $Q_\mathrm{r}$ and only at higher frequencies damping of the Bessel-type SPP prevails and limits the quality factor.
The predictions for the quality factor were used to calculate the peak widths $\Delta \nu$. They agree very well with those seen in Fig.~\ref{fig:FreqScan}~a) for the numerically simulated response.

\subsection{\label{sub:Stack_System_Study}A Simple Stack System}

Until now for the sake of simplicity the theory was tested for an isolated metallic disc. However, the theory is also applicable for stacked nanoantennas as shown in Fig.~\ref{fig:axialStack}. Such stacks introduce a much larger degree of freedom for tailoring the properties of Bessel-type SPPs concerning both their propagation constant and their complex reflection coefficient at the disc termination. In this subsection, it will be shown that the theory is indeed applicable to these kinds of systems by verifying it for a selected case.

\begin{figure}
\begin{centering}
\includegraphics[width=80mm]{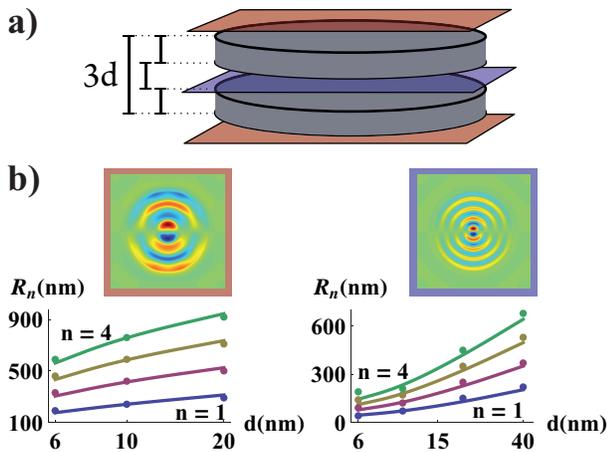}
\par\end{centering}
\caption{\label{fig:twoDiscResult}
a) Schematic of the stacked nanoantenna - two silver discs with an air cavity in-between. Every layer has a thickness $d$.
b) For a non-resonant radius of $R=810$~nm and thickness $d=20$~nm, $E_z$ is plotted for the symmetric and antisymmetric modes directly from simulations. The field profile for the symmetric mode on the left (orange frame) was computed $20$~nm above the structure, whereas the antisymmetric one (blue frame) directly in the middle.
Also, a comparison of the theoretical predictions (full lines) against simulation results (dots) is shown. Again, a very good agreement is found.
It has to be mentioned that for a thickness of $d=40$~nm the resonances of the symmetric mode could not be identified. There, the antisymmetric mode turns out to be dominating.
}
\end{figure}

Namely, the nanoantenna under consideration shall consist of two silver discs, each of thickness $d$. They are separated by a dielectric spacer of the same thickness. For reasons of simplicity air was assumed as the spacer and the surrounding medium. The configuration is illustrated in Fig.~\ref{fig:twoDiscResult}~a).

Generally, the structure supports several Bessel-type plasmonic modes.
However, only two modes can be excited efficiently. They can be understood as the symmetric and antisymmetric coupling of the metallic discs.
Thus, the modes are termed symmetric and antisymmetric in correspondence to the symmetry class of $E_\rho$ with respect to the $z$-coordinate. If $E_\rho$ has a certain symmetry, $E_z$ belongs to the opposite symmetry class - $E_z$ is symmetric if $E_\rho$ is antisymmetric and vice versa. Hence, $E_z$ vanishes for the symmetric mode in the center of the air cavity but is strongest there for the antisymmetric one. Furthermore, the symmetric mode is confined mostly at the lower and upper terminations of the structure. This spatial separation of the modes make it possible to identify resonances of the nanoantenna corresponding to the different modes. More on the nomenclature, dispersion relation and profile of the modes can be found in App.~\ref{sub:twoDiscDispRelation}.

The simulations were analogue to the ones outlined in subsection~\ref{sub:Verification_Bessel_and_Scaling}. The nanoantennas were illuminated by a plane wave propagating in $z$-direction at a constant frequency $\nu=625$~THz. There, the permittivity of silver\cite{Johnson1972} is $\epsilon_{m}\approx-8.8+0.3\mathrm{i}$.
The radii $R$ of the antennas varied from $40$~nm up to $1000$~nm and $d$ was chosen to range from $6$~nm to $40$~nm. Absolute values of $E_z$ were monitored in the middle of the structure and $20$~nm above and below to identify resonances of the odd and even modes, respectively. The results are outlined in Fig.~\ref{fig:twoDiscResult} b).

%%%
Hence, it may be concluded that the reflection coefficient as calculated from Eq.~\ref{eq:reflectionCoeff_most_general} can be
used to evaluate the resonant behavior for the given class of nanoantennas if the dominant modes inside the resonator are the assumed Bessel-type SPPs.

\section{Conclusion}

A theory to describe radially propagating Hankel-type SPPs in piecewise homogeneous circular nanoresonators was introduced.
The complex reflection coefficients at the termination were analytically calculated, thus yielding the phase change of the
SPPs upon reflection. In combination with a Fabry-Perot model that predicts the resonances of circular
patch nanoantennas, the theory was proven to be valid while comparing results to full-wave simulations.

This theory constitutes a unique approach towards the analytical discussion of the resonant properties of circular optical nanoantennas without requiring any fitting parameters. This allows for a deeper insight into the scaling behavior and will foster further research since a desired simple tool is now available to design such nanoantennas with only little effort. It can be anticipated, for example, that the theory can be applied to design antennas supporting various resonances at predefined frequencies to respond on the desire to have multiresonant antennas at hand for applications in, e.g., Raman-sensors or extremely broad-band resonators. The unique design tool of such circular patch nanoantennas consists in tailoring the dispersion relation and the complex reflection coefficient at will by carefully selecting a particular stack of layers. This provides a large degree of freedom which renders such optical nanoantennas unique.

\section*{Acknowledgements}

Financial support by the German Federal Ministry of Education and Research (PhoNa), by the Thuringian State Government (MeMa) and the German Science
Foundation (SPP 1391 Ultrafast Nano-optics) is gratefully acknowledged. RF thanks F. Maucher for encouraging discussions. Furthermore we acknowledge the substantial suggestions of the anonymous referees. We also thank K. Verch (www.karstenverch.com) for providing us with the artist's view of the theory as shown in Fig.~\ref{fig:ToC}.

%\begin{widetext}
\onecolumngrid
\section*{Appendix}
\appendix

\section{\label{sec:Derivation_of_the_Refl_Coeff}Derivation of the Reflection
Coefficient}

In this section, the complex reflection coefficient for Hankel-type
surface plasmon polaritons (SPPs) for the given class of nanoantennas will be derived. For a definition of the geometry see again Fig.~\ref{fig:axialStack}. At first, radially propagating
plasmonic fields inside the resonator and a suitable ansatz for the outgoing field will be examined. At second, by calculating the Fourier components
of the radiating field and by using the continuity of $H_{\varphi}^{m}$,
one can finally calculate the reflection coefficient
integrating $E_{z}^{m}\cdot H_{\varphi}^{m}$ at $\rho=R$.

\subsection{Plasmonic Modes and Outer Fields}

The studied nanoantennas obey a piecewise translational symmetry in $z$-direction. Thus, the structure consists of several layers which may be denoted by the subscript $i$ in the following.
It will be assumed that the antenna supports plasmonic surface modes propagating in radial direction. Hence, at least one of the materials is a metal.

In each layer, one may split the fields into tangential ($\mathbf{E}_{i,t},\ \mathbf{H}_{i,t}$) and normal components ($E_{i,z}, H_{i,z}$) relative to the interface of the discs. Then, the entire dynamic can be calculated using only the normal components. The tangential fields in the $i$-th layer are given by
\begin{eqnarray}
\mathbf{E}_{i,t} & = & \frac{\mathrm{i}}{k_{\rho}^{2}}\left[k_{i,z}\nabla_{t}E_{i,z}-\omega\mu_{0}\mu_{i}\mathbf{e}_{z}\times\nabla_{t}H_{i,z}\right]\ \mathrm{and}\nonumber \\
\mathbf{H}_{i,t} & = & \frac{\mathrm{i}}{k_{\rho}^{2}}\left[k_{i,z}\nabla_{t}H_{i,z}+\omega\epsilon_{0}\epsilon_{i}\mathbf{e}_{z}\times\nabla_{t}E_{i,z}\right]\label{eq:Decomposition_Et_and_Ht}
\end{eqnarray}
with $k_{i,z}$ being the wave vector component in $z$-direction. In this formulation, all fields obey $\exp\left(\mathrm{i}k_{i,z}z\right)$ and $\exp\left(-\mathrm{i}\omega t\right)$ dependencies\cite{Jackson98}.

The radial wave vector
\begin{eqnarray*}
k_{\rho}^{2} & = & \frac{\omega^{2}}{c^{2}}\mu_{i}\epsilon_{i}-k_{i,z}^{2}
\end{eqnarray*}
has to be independent of $z$ since it is a conserved quantity in all layers.
The various plasmonic modes of the antenna are thus characterized by different $k_{\rho}$ and it is convenient to define both
\begin{eqnarray*}
\gamma_{i} & = & \mathrm{i}k_{i,z}\ \mathrm{and}\\
k_{\mathrm{SPP}} & \equiv & k_{\rho}\ .
\end{eqnarray*}
In general, the structure may support several modes. Nevertheless, for the sake of simplicity, a further corresponding index will be suppressed.

Furthermore, transverse-magnetic (TM) modes with $H_{z}=0$ will be assumed and the materials shall be non-magnetic, i.e. $\mu_{i}=1$. Thus,
\begin{eqnarray}
H_{z} & = & 0\ ,\ \mathrm{and}\nonumber \\
\mathbf{H}_{i,t} & = & \frac{\mathrm{i}}{k_{\mathrm{SPP}}^{2}}\omega\epsilon_{0}\epsilon_{i}\mathbf{e}_{z}\times\nabla_{t}E_{i,z}\nonumber \\
& = & \frac{\mathrm{i}}{k_{\mathrm{SPP}}^{2}}\omega\epsilon_{0}\epsilon_{i}\mathbf{e}_{z}\times
\left[\partial_{\rho}E_{i,z}\mathbf{e}_{\rho}+\frac{1}{\rho}\left(\partial_{\varphi}E_{i,z}\right)\mathbf{e}_{\varphi}\right]\ .\label{eq:H_transverse}
\end{eqnarray}
The only nonvanishing normal field component $E_z$ obeys
the scalar two-dimensional Helmholtz equation\cite{Jackson98}
\begin{eqnarray*}
\Delta_t E_{z}+k_{\mathrm{SPP}}^{2}E_{z} & = & 0\ .
\end{eqnarray*}
Since $k_{\mathrm{SPP}}$ is a constant throughout the layers, this differential equation generally applies to the corresponding plasmonic mode and the index $i$ can be dropped.
In rotational symmetry, the latter equation is known to have solutions in terms of Bessel and Neumann functions $J_{m}\left(k_{\mathrm{SPP}}\rho\right)$ and $N_{m}\left(k_{\mathrm{SPP}}\rho\right)$ modulus an angular variation of $\exp{\left(\mathrm{i}m\varphi\right)}$ which will be denoted by the index $m$ in the following.

For the calculation of the complex reflection coefficient one has to use
Hankel functions which describe propagating rather than standing wave fields. They read as
\begin{eqnarray*}
H_{m}^{1/2}\left(k_{\mathrm{SPP}}\rho\right) & \equiv & J_{m}\left(k_{\mathrm{SPP}}\rho\right)\pm\mathrm{i}N_{m}\left(k_{\mathrm{SPP}}\rho\right)
\end{eqnarray*}
and represent radially outgoing (1) and incoming (2) cylindrical waves in the given time-dependence.
Since the fields are given in terms of $E_{z}$ it is convenient
to introduce the reflection coefficient $r_{m}$ as
\begin{eqnarray*}
E_{z}^{m,-}\left(\rho,z\right) & = & \mathcal{A}_{m}\left(k_{\mathrm{SPP}}\rho\right)\cdot a\left(z\right)\ \mathrm{with}\\
\mathcal{A}_{m}\left(k_{\mathrm{SPP}}\rho\right) & = & H_{m}^{1}\left(k_{\mathrm{SPP}}\rho\right)+r_{m}\cdot H_{m}^{2}\left(k_{\mathrm{SPP}}\rho\right)
\end{eqnarray*}
where the boundary conditions at the antennas termination determine the {}``mode
profile'' $a\left(z\right)$ in $z$-direction as well as the propagation constant $k_{\mathrm{SPP}}$. Furthermore, the superscript $-$ was used to indicate fields for $\rho\leq R$. Likewise, $+$ will be used in the outer region.

From Eq.~\ref{eq:H_transverse} the angular component of the magnetic field is given by
\begin{eqnarray}
H_{\varphi}^{m,-}\left(\rho,z\right) & = & \frac{\mathrm{i}\omega\epsilon_{0}\epsilon_{i}}{\frac{\omega^{2}}{c^{2}}\epsilon_{i}+\gamma_{i}^{2}}\partial_{\rho}E_{z}^{m,-}\left(\rho,z\right)\label{eq:Helmholtz_piecewise}\\
 & = & \frac{\mathrm{i}\omega\epsilon_{0}}{k_{SPP}}\mathcal{B}_{m}\left(k_{SPP}\rho\right)\epsilon\left(z\right)a\left(z\right)\nonumber
 \end{eqnarray}
with
\begin{eqnarray*}
\mathcal{B}_{m}\left(k_{SPP}\rho\right) & = & \frac{1}{k_{SPP}}\partial_{\rho}\mathcal{A}_{m}\left(k_{SPP}\rho\right)\\
 & = & DH_{m}^{1}\left(k_{SPP}\rho\right)+r_{m}\cdot DH_{m}^{2}\left(k_{SPP}\rho\right)\ ,\\
DH_{m}^{1/2}\left(x\right) & = & \partial_{x}H_{m}^{1/2}\left(x\right)\ .
\end{eqnarray*}

Having determined $E_{z}^{m,-}$ and $H_{\varphi}^{m,-}$,
a reasonable ansatz for the fields outside the antenna exhibiting the same
symmetry properties, i.e. the same angular modulus $m$, can be constructed. For $\rho\geq R$, both fields
under consideration are a superposition
of outgoing Hankel-type SPPs whose amplitudes are given by certain
Fourier coefficients $c_{m}\left(k_{z}\right)$.
For each of these waves the wave vector in radial direction is given by
\begin{eqnarray*}
k_{\rho}^{+} & = & \sqrt{\epsilon_{d}k_{0}^{2}-k_{z}^{2}}\ .
\end{eqnarray*}
Here, a relative permittivity $\epsilon_{d}$ is assumed in the outer region. With this ansatz, the fields
can be represented as
\begin{eqnarray}
E_{z}^{m,+}\left(\rho,z\right) & = & \int_{-\infty}^{\infty}c_{m}\left(k_{z}\right)H_{m}^{1}\left(\sqrt{\epsilon_{d}k_{0}^{2}-k_{z}^{2}}\rho\right)e^{\mathrm{i}k_{z}z}dk_{z}\ ,\nonumber \\
H_{\varphi}^{m,+}\left(\rho,z\right) & = & \mathrm{i}\int_{-\infty}^{\infty}c_{m}\left(k_{z}\right)\frac{\epsilon_{0}\epsilon_{d}\omega}{\sqrt{\epsilon_{d}k_{0}^{2}-k_{z}^{2}}}\\
& & \cdot DH_{m}^{1}\left(\sqrt{\epsilon_{d}k_{0}^{2}-k_{z}^{2}}\rho\right) e^{\mathrm{i}k_{z}z}dk_{z}\label{eq:decomposition_outside}
\end{eqnarray}
where the magnetic field follows from Eq.~\ref{eq:H_transverse}.

\subsection{The Fourier Components of the Radiating Field}

In this section, the continuity of the magnetic field will be
used to determine the Fourier components $c_{m}\left(k_{z}\right)$.
Since $\mathcal{B}_m$, by definition, explicitly depend on $r_m$, this will likewise hold for the $c_m\left(k_{z}\right)$.

A Fourier transformation for $\rho\leq R$ yields
\begin{eqnarray*}
\int_{-\infty}^{\infty}H_{\varphi}^{m,-}\left(\rho,z\right)e^{-\mathrm{i}k_{z}z}dz & = & \frac{\mathrm{i}\omega\epsilon_{0}}{k_{SPP}}\mathcal{B}_{m}\left(k_{SPP}\rho\right)\cdot B^{-}\left(k_{z}\right)\ \\
\mathrm{with}\  B^{\pm}\left(k_{z}\right) & \equiv & \int_{-\infty}^{\infty}\epsilon\left(z\right)a\left(z\right)e^{\pm\mathrm{i}k_{z}z}dz\ .
\end{eqnarray*}

The same operation for the outer region results in
\begin{eqnarray*}
\int_{-\infty}^{\infty}H_{\varphi}^{m,+}\left(\rho,z\right)e^{-\mathrm{i}k_{z}z}dz & = & \mathrm{i}\int_{-\infty}^{\infty}c_{m}\left(k\right)
\frac{\epsilon_{0}\epsilon_{d}\omega}{\sqrt{\epsilon_{d}k_{0}^{2}-k^{2}}} DH_{m}^{1}\left(\sqrt{\epsilon_{d}k_{0}^{2}-k^{2}}\rho\right) \underbrace{\int_{-\infty}^{\infty}e^{\mathrm{i}\left(k-k_{z}\right)z}dz}
_{2\pi\delta\left(k-k_{z}\right)}dk\\
& = & 2\pi\mathrm{i}\cdot c_{m}\left(k_{z}\right)
\frac{\epsilon_{0}\epsilon_{d}\omega}{\sqrt{\epsilon_{d}k_{0}^{2}-k_{z}^{2}}}\cdot DH_{m}^{1}\left(\sqrt{\epsilon_{d}k_{0}^{2}-k_{z}^{2}}\rho\right)\ .\end{eqnarray*}
Matching at $\rho=R$ gives
\begin{eqnarray*}
\frac{\mathrm{i}\omega\epsilon_{0}}{k_{SPP}}\mathcal{B}_{m}\left(k_{SPP}\rho\right)B^{-}\left(k_{z}\right) & = & 2\pi\mathrm{i}\cdot c_{m}\left(k_{z}\right)\frac{\epsilon_{0}\epsilon_{d}\omega}{\sqrt{\epsilon_{d}k_{0}^{2}-k_{z}^{2}}}\\ & & \cdot DH_{m}^{1}\left(\sqrt{\epsilon_{d}k_{0}^{2}-k_{z}^{2}}R\right)
\end{eqnarray*}
further leading to the desired
\begin{eqnarray}
c_{m}\left(k_{z}\right) & = & \frac{\frac{\mathrm{i}\omega\epsilon_{0}}{k_{SPP}}\mathcal{B}_{m}\left(k_{SPP}R\right)\cdot B^{-}\left(k_{z}\right)}{2\pi\mathrm{i}\cdot\frac{\epsilon_{0}\epsilon_{d}\omega}{\sqrt{\epsilon_{d}k_{0}^{2}-k_{z}^{2}}}DH_{m}^{1}\left(\sqrt{\epsilon_{d}k_{0}^{2}-k_{z}^{2}}R\right)}\nonumber \\
& = & \frac{\mathcal{B}_{m}\left(k_{SPP}R\right)}{2\pi\cdot\epsilon_{d}\cdot k_{SPP}}\cdot\frac{\sqrt{\epsilon_{d}k_{0}^{2}-k_{z}^{2}}B^{-}\left(k_{z}\right)}{DH_{m}^{1}\left(\sqrt{\epsilon_{d}k_{0}^{2}-k_{z}^{2}}R\right)}\ .\label{eq:Hankelgeneral_cm}
\end{eqnarray}

\subsection{Integration over $E_{z}^{m}\cdot H_{\varphi}^{m}$}

The final step to calculate the reflection coefficient is the integration
of the inner and outer $E_{z}^{m}$ using $H_{\varphi}^{m,-}$ at
$\rho=R$. Since there the factor $\mathcal{B}_{m}\left(k_{SPP}R\right)$
is a constant, it is sufficient to use the field profile $\epsilon\left(z\right)\cdot a\left(z\right)$
alone.

First, one finds
\begin{eqnarray*}
\int_{-\infty}^{\infty}E_{z}^{m,-}\left(R,z\right)\epsilon\left(z\right)a\left(z\right)dz
 & = & \mathcal{A}_{m}\left(k_{SPP}R\right)\cdot\sigma\ \mathrm{with}\\
 \sigma & = & \int_{-\infty}^{\infty}a\left(z\right)^2\cdot\epsilon\left(z\right)dz\ .\end{eqnarray*}

Performing the same integration over the mode field for $\rho\geq R$ gives
\begin{eqnarray*}
\int_{-\infty}^{\infty}E_{z}^{m,+}\left(R,z\right)\epsilon\left(z\right)a\left(z\right)dz
 & = & \int_{-\infty}^{\infty}c_{m}\left(k_{z}\right)H_{m}^{1}\left(\sqrt{\epsilon_{d}k_{0}^{2}-k_{z}^{2}}R\right) \cdot\left(\int_{-\infty}^{\infty}e^{\mathrm{i}k_{z}z}\epsilon\left(z\right)a\left(z\right)dz\right)dk_{z} \\
 & = & \int_{-\infty}^{\infty}c_{m}\left(k_{z}\right)H_{m}^{1}\left(\sqrt{\epsilon_{d}k_{0}^{2}-k_{z}^{2}}R\right) \cdot B^{+}\left(k_{z}\right)dk_{z}\ .
\end{eqnarray*}
and one can equate the last two results obtaining
\begin{eqnarray*}
\mathcal{A}_{m}\left(k_{SPP}R\right)\cdot\sigma & = & \int_{-\infty}^{\infty}c_{m}\left(k_{z}\right)H_{m}^{1}\left(\sqrt{\epsilon_{d}k_{0}^{2}-k_{z}^{2}}R\right) \\ &  & \cdot B^{+}\left(k_{z}\right)dk_{z}\ .
\end{eqnarray*}
Further using Eq.~\ref{eq:Hankelgeneral_cm} with the known  $c_{m}\left(k_{z}\right)$,
one finds\begin{eqnarray}
\mathcal{A}_{m}\left(k_{SPP}R\right)\cdot\sigma
 & = & \frac{\mathcal{B}_{m}\left(k_{SPP}R\right)}{2\pi\cdot\epsilon_{d}\cdot k_{SPP}}\cdot I_{m}\label{eq:Reflection_general_Prestep}\end{eqnarray}
 with \begin{eqnarray}
 I_{m} & = & \int_{-\infty}^{\infty}\frac{H_{m}^{1}\left(\sqrt{\epsilon_{d}k_{0}^{2}-k_{z}^{2}}R\right)}{DH_{m}^{1}\left(\sqrt{\epsilon_{d}k_{0}^{2}-k_{z}^{2}}R\right)}\nonumber \\
 &  & \cdot \sqrt{\epsilon_{d}k_{0}^{2}-k_{z}^{2}}B^{-}\left(k_{z}\right) \cdot B^{+}\left(k_{z}\right)dk_{z}\ .\nonumber \end{eqnarray}

Finally inserting the definitions of $\mathcal{A}_{m}$ and $\mathcal{B}_{m}$, one can calculate the reflection coefficient as it is given in Eq.~\ref{eq:reflectionCoeff_most_general} in the main text,
\begin{eqnarray*}
r_{m} & = & \frac{2\pi\cdot\epsilon_{d}\cdot k_{SPP}\sigma\cdot H_{m}^{1}\left(k_{SPP}R\right)-DH_{m}^{1}\left(k_{SPP}R\right)\cdot I_{m}}{-2\pi\cdot\epsilon_{d}\cdot k_{SPP}\sigma\cdot H_{m}^{2}\left(k_{SPP}R\right)+DH_{m}^{2}\left(k_{SPP}R\right)\cdot I_{m}}\ .
\end{eqnarray*}
Also note that every rescaling in $a\left(z\right)$ does not affect this result due to the linearity of the underlying theory.

\section{\label{sec:Comparison_to_Previous_Results}Comparison to Previous Results}

It is known that Hankel-type SPPs have the same dispersion relation as surface
plasmons at planar surfaces\cite{Nerkararyan2010} which can be understood from the expansion
of the Hankel functions for large arguments.

The present theory can be applied to calculate the reflection coefficient of Hankel-type SPPs on top of a semi-infinite cylinder with the dielectric function $\epsilon_{m}(\omega)$.
The dispersion relation of a SPP at such an interface between a dielectric and a metal is then simply given by\cite{Sernelius2001,Maier2007}
\begin{eqnarray}
k_{\mathrm{SPP}}(\omega) & = & \frac{\omega}{c} \sqrt{\frac{\epsilon_{m}(\omega)\epsilon_{d}}{\epsilon_{m}(\omega)+\epsilon_{d}}}\ .\label{eq:DispRelation_semiInfinite}
\end{eqnarray}
Since the expression is explicit, $r_{m}$ can also be explicitly derived, see App.~\ref{sec:Half-Infinite-cylinder-rm}.

\begin{figure}
\begin{centering}
\includegraphics[width=160mm]{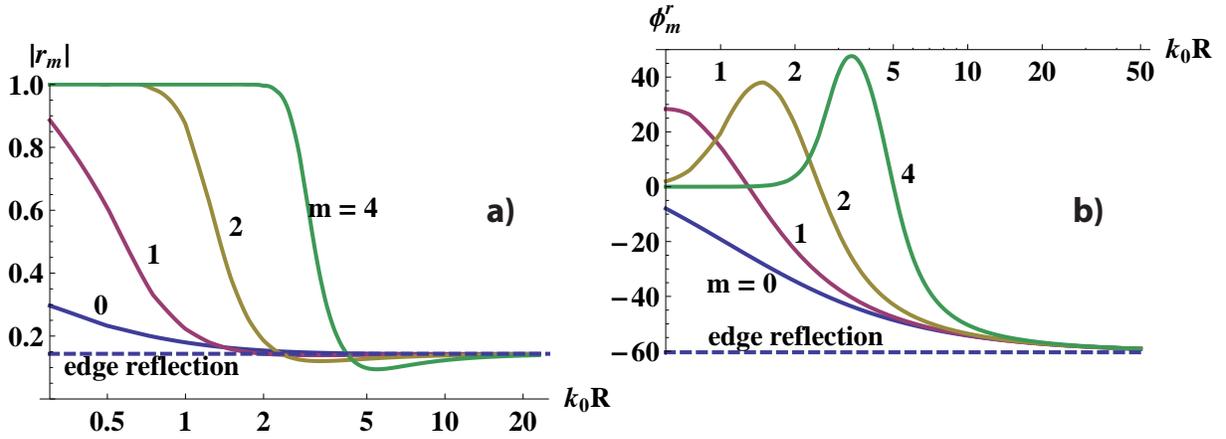}
\par\end{centering}
\caption{\label{fig:Comparison_reflection_hankel_and_edge-1}
The amplitude a) and phase b) of the reflection coefficient for different angular mode numbers $m$ as a function of the radius $R$ (solid lines). Assumed is a metal with a permittivity of  $\epsilon_{m}=-15+\mathrm{i}0^{+}$ surrounded by air, i.e. $\epsilon_{d}=1$. Results are compared to previous expressions that have been obtained for planar interfaces\cite{Gordon2006} at normal incidence (dashed lines).}
\end{figure}

A further limiting case concerns the reflection of SPPs at an ordinary planar interface which is obtained for an infinite radius $R\rightarrow\infty$.
The result can be used to verify the theory in this limit since analytical expressions for such a specific geometry are available \cite{Gordon2006}.
Results are outlined in Fig.~\ref{fig:Comparison_reflection_hankel_and_edge-1}. There, both the modulus and the phase of the reflection coefficient
are shown as a function of the radius of the semi-infinite cylinder for modes with different angular mode number $m$. It can be seen that the results
as predicted by this theory converge for an infinite radius nicely towards the results previously reported\cite{Gordon2006}.

Moreover, the theory discloses that there exists an intermediate region sustaining a very complex dependency with the emergence of a resonance behavior. It is associated with an abrupt decrease of the modulus and the emergence of a peak in the phase of the reflection coefficient. This region narrows for higher angular modes indicating a steep transition that might be of great practical use in the future, i.e. for a minor radius change the phase of the reflection coefficient changes abruptly; suggesting it for sensor applications.

At a fixed radius, the reflection coefficient for discs of increasing thickness should converge to those of the semi-infinite disc. It is important to verify this point not only to check the consistency of the theory but also to see if one can approximately use the reflection coefficient of the semi-infinite case.

\begin{figure}
\begin{centering}
\includegraphics[width=160mm]{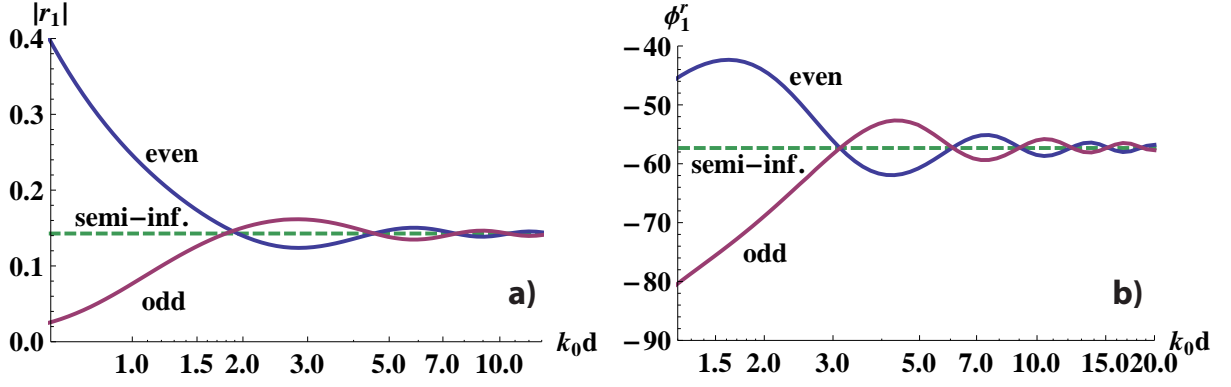}
\par\end{centering}
\caption{\label{fig:Infinite-to-slab-compare}Comparison of $r_{1}$ for $\epsilon_{m}=-15+\mathrm{i}0^{+}$
and $\epsilon_{d}=1$ with the semi-infinite cylinder model at $k_{0}R=20$
which is approached as the thickness $d$ goes to infinity both in absolute value and phase, see a) and b).}
\end{figure}

Figure~\ref{fig:Infinite-to-slab-compare} shows that for $d\rightarrow\infty$,
the dipolar reflection coefficients of the even and odd modes approach
that of the semi-infinite approximation for a radius of $k_{0}R=20$
and $\epsilon_{m}=-15+\mathrm{i}0^{+}$, $\epsilon_{d}=1$.

The absolute value of $r_m$ converges more rapidly in the given case. However, one can see that already at $k_0\cdot d \gtrsim 3$, the values do not vary
strongly with respect to the semi-infinite value which would justify to use this approximation here.

From all these considerations it may be concluded that in limiting cases the theory reproduces previous results;
being an indication for its performance. It can be also seen that although the theory might have restrictiions
in its formulation; it turns out to be rather general since many other geometries can be deduced from it while
considering the limits of the theory.

\section{\label{sec:Explicit-disc-modes}Explicit Plasmonic Modes}

It was shown \cite{Nerkararyan2010} that a Hankel-type
plasmonic mode at a dielectric-metal surface has the same dispersion
relation as the one known for a plane-wave plasmonic mode.
This is the general case as can
be seen by the infinite-radius approximations of the cylindrical Hankel
functions. In this section, the analytic expressions for plasmonic
modes as used in the main body of the manuscript will be explicitly given for convenience -
for semi-infinite cylinders, discs with finite thicknesses and a
stacked antenna consisting of two metallic discs separated by a dielectric spacer.

\subsection{\label{sub:Semi-Infinite-cylinder}Semi-Infinite Cylinder}

Assumed is a cylinder of radius $R$ terminated at $z=0$. Its relative permittivity is then described by
\begin{eqnarray*}
\epsilon\left(z\right) & = & \begin{cases}
\epsilon_{d} & z>0\\
\epsilon_{m} & z<0\end{cases}\end{eqnarray*}
where
\begin{eqnarray*}
\epsilon_{d}^{\prime} & \geq & 0\ \mathrm{and}\\
\epsilon_{m}^{\prime} & < & -\epsilon_{d}^{\prime}\ .
\end{eqnarray*}

In this case, a Hankel-type surface plasmon polariton exists
with the mode profile \begin{eqnarray*}
a\left(z\right) & = & \begin{cases}
\epsilon_{m}e^{-\gamma_{d}z} & z\geq0\\
\epsilon_{d}e^{\gamma_{m}z} & z\leq0\end{cases}\ ,\\
\epsilon\left(z\right)a\left(z\right) & = & \epsilon_{d}\epsilon_{m}\begin{cases}
e^{-\gamma_{d}z} & z\geq0\\
e^{\gamma_{m}z} & z\leq0\end{cases}\end{eqnarray*}
and the well-known dispersion relation \cite{Sernelius2001,Maier2007}
\begin{eqnarray}
k_{SPP}^{2} & = & \frac{\epsilon_{m}\epsilon_{d}}{\epsilon_{m}+\epsilon_{d}}k_{0}^{2}\ \mathrm{and}\nonumber \\
\gamma_{i}^{2} & = & k_{SPP}^{2}-\epsilon_{i}k_{0}^{2}\nonumber \\
& = & -\frac{\epsilon_{i}^{2}}{\epsilon_{d}+\epsilon_{m}}k_{0}^{2}\ .\label{eq:Dispersion_semi_infinite}
\end{eqnarray}
By definition the $\gamma_{d/m}$ are related to penetration
depths of the plasmonic fields $d_{p,m/d}=\left(\gamma_{m/d}^{\prime}\right)^{-1}$.

\subsection{\label{sub:Disc-with-finite-thickness}Disc with Finite Thickness}

In this subsection, the case of a metallic disc with radius $R$ surrounded by another dielectric material will be considered. Then, for $\rho<R$, the relative permittivity is given by
\begin{eqnarray*}
\epsilon\left(z\right) & = & \begin{cases}
\epsilon_{d} & z>d/2\\
\epsilon_{m} & \left|z\right|\leq d/2\\
\epsilon_{d} & z<-d/2\end{cases}\ .\end{eqnarray*}

In this case, there are two transverse magnetic modes supported by
the system. They can be described by
\begin{eqnarray*}
a_{\pm}\left(z\right) & = & \begin{cases}
\epsilon_{m}\exp\left[-\gamma_{d}\left(z-\frac{d}{2}\right)\right] & z>\frac{d}{2}\\
\mp\epsilon_{m}\exp\left[\gamma_{d}\left(z+\frac{d}{2}\right)\right] & z<\frac{d}{2}\\
\epsilon_{d}\frac{\text{hyp}_{\mp}\left(\gamma_{m}z\right)}{\text{hyp}_{\mp}\left(\gamma_{m}\frac{d}{2}\right)} & \left|z\right|\leq\frac{d}{2}\end{cases}\ ,\end{eqnarray*}
and the dispersion relation is given by \begin{eqnarray*}
\tanh\left(\gamma_{m}\frac{d}{2}\right) & = & -\left[\frac{\gamma_{m}\epsilon_{d}}{\gamma_{d}\epsilon_{m}}\right]^{\pm1}\ \mathrm{with}\\
\gamma_{i}^{2} & = & k_{SPP}^{2}-\epsilon_{i}k_{0}^{2}\ .\end{eqnarray*}

Modes which are odd in $E_{z}$, thus odd in $a\left(z\right)$ are
called even \cite{Maier2007} and denoted here by $a_{+}$ and $b_{+}$. In this case,
$\text{hyp}_{-}\left(x\right)\equiv\sinh\left(x\right)$. An illustration of this mode can be found in Fig.~\ref{fig:ModeProfiles}~a).
For odd modes, the corresponding conventions hold along with $\text{hyp}_{+}\left(x\right)\equiv\cosh\left(x\right)$.
Then, one finds $\sigma$ and $B^{-}\left(k\right)\cdot B^{+}\left(k\right)$
by direct computation.

\subsection{\label{sub:twoDiscDispRelation}Stacks with Two Metallic Discs}

Assumed is that two metallic discs with radius $R$ and thickness $d_m$ are separated by some dielectric with thickness $d_d$. Furthermore, this simple stack is surrounded by the same dielectric, hence
\begin{eqnarray*}
\epsilon\left(z\right) & = & \begin{cases}
\epsilon_{d} & z<0,\ d_m<z<d_m + d_d\ \mathrm{and}\ 2d_m + d_d<z\ ,\\
\epsilon_{m} & 0\leq z\leq d_m\ \mathrm{and}\ d_m + d_d\leq z\leq 2d_m + d_d
\end{cases}\ .\end{eqnarray*}

The structure supports several Bessel-type plasmonic modes that can be characterized by
\begin{eqnarray*}
a_{i}\left(z\right) & = &
\begin{cases}
\epsilon_m C_{1+,i}e^{\gamma_{d,i} z} & z<0\\
\epsilon_d C_{2+,i}e^{\gamma_{m,i} (z-d_m)} + \epsilon_d C_{2-,i}e^{-\gamma_{m,i} z} & 0\leq z\leq d_m\\
\epsilon_m C_{3+,i}e^{\gamma_{d,i} (z-d_m-d_d)} + \epsilon_m C_{3-,i}e^{-\gamma_{d,i} (z-d_m)} & d_m<z<d_m + d_d\\
\epsilon_d C_{4+,i}e^{\gamma_{m,i} (z-2d_m-d_d)} + \epsilon_d C_{4-,i}e^{-\gamma_{m,i} (z-d_m-d_d)} & d_m + d_d\leq z\leq 2d_m + d_d\\
\epsilon_m C_{5-,i}e^{-\gamma_{d,i} (z-2d_m-d_d)} & 2d_m + d_d<z
\end{cases}\ ,\end{eqnarray*}
with $\gamma_{m/d,i}^{2} = k_{SPP,i}^{2}-\epsilon_{m/d,i}k_{0}^{2}$.
The dispersion relation and the coefficients $C_{l\pm,i}$ can be found by applying the transfer matrix method\cite{Sernelius2001} which can also be used for more complex stack compositions.

\begin{figure}
\begin{centering}
\includegraphics[width=160mm]{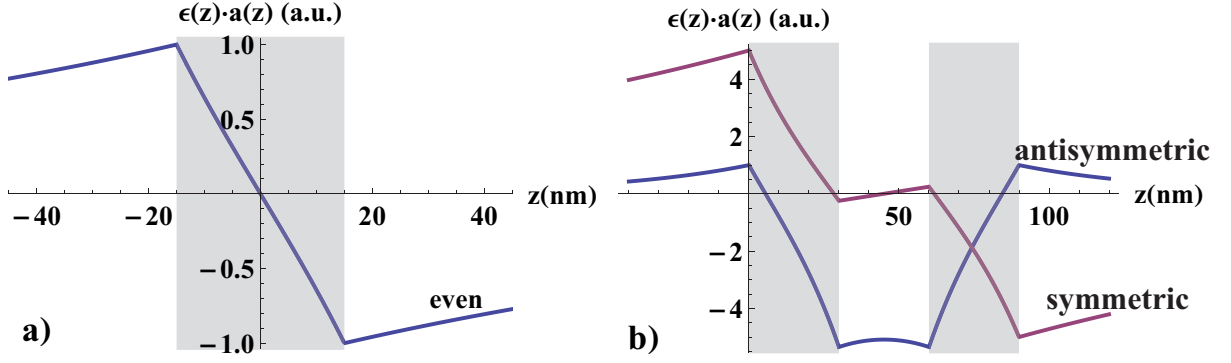}
\par\end{centering}
\caption{\label{fig:ModeProfiles}
Modes that could be observed in the numerical experiments performed. The material of the metals is taken to be silver with $\epsilon_m\approx-8.8+0.3\mathrm{i}$ at $\nu=625$~THz.
In a), one can see $\epsilon\left( z \right)a\left( z \right)$ for the even mode of a single disc with $30$~nm thickness. The same quantity is outlined in b) for the symmetric and antisymmetric modes of a nanoantenna consisting of two silver discs with $30$~nm thickness and also $30$~nm air cavity in between. These modes can be understood as symmetric and antisymmetric coupling between single discs obeying an even mode. One can directly see that in this case the antisymmetric mode is strongly confined inside the cavity. On the contrary, $E_z$ of the symmetric mode is mostly located outside the metal discs.}
\end{figure}

In the study of Section \ref{sub:Stack_System_Study}, a symmetric situation was used with $d_d=d_m$ such that only two dominating modes have to be considered.
The first one is characterized by
$$C_{1+,1} = C_{5-,1},\ C_{3+,1}=C_{3-,1},\ C_{2-,1} = C_{4+,1}\ \mathrm{and}\ C_{2+,1} = C_{4-,1}\ .$$
With the explanations given in the preceding subsection, it is natural to call this mode antisymmetric since it is symmetric in $E_z\propto a(z)$. The second mode is given due to the relations
$$C_{1+,2} = -C_{5-,2},\ C_{3+,2}=-C_{3-,2},\ C_{2-,2} = -C_{4+,2}\ \mathrm{and}\ C_{2+,2} = -C_{4-,2}$$ and consequentially termed symmetric. The modes are illustrated in Fig.~\ref{fig:ModeProfiles}~b); see again Fig.~\ref{fig:twoDiscResult} for an illustration of the nanoantenna.
However, the explicit coefficients $C_{l\pm,i}$ and corresponding dispersion relations for $k_{S\!P\!P,i}$ turn out to be rather complicated to be presented here.

\section{\label{sec:Half-Infinite-cylinder-rm}Explicit Reflection Coefficient
for a Semi-Infinite Cylinder}

The underlying physical situation of SPPs on a semi-infinite cylinder
is a limit case in two aspects: For large radii, the reflection coefficient
is converging towards that of SPPs for normal incidence at a planar interface\cite{Gordon2006}.
Furthermore, it is equivalent to the reflection of Hankel-type
SPPs on cylinders of large thicknesses. This relation was used in
the main body of the manuscript to verify the calculations.

In the given case, the dispersion relation takes an explicit form,
see Eq.~\ref{eq:Dispersion_semi_infinite}, and it is possible
to find a more explicit formula for the reflection coefficient which
will be derived in the following.

As a preliminary step, one has to calculate $B^{-}\left(k\right)\cdot B^{+}\left(k\right)$ and $\sigma$ using the permittivity $\epsilon\left(z\right)$ and mode profile $a\left(z\right)$
\begin{eqnarray*}
B^{-}\left(k\right)\cdot B^{+}\left(k\right) & = & \epsilon_{d}^{2}\epsilon_{m}^{2}\frac{\left(\gamma_{d}+\gamma_{m}\right)^{2}}{\left(k^{2}+\gamma_{d}^{2}\right)\left(k^{2}+\gamma_{m}^{2}\right)}\ ,\\
\sigma  & = & \frac{\epsilon_{d}\epsilon_{m}}{2\gamma_{d}\gamma_{m}}\left(\gamma_{d}\epsilon_{d}+\gamma_{m}\epsilon_{m}\right)\\
 & = & \frac{\epsilon_{d}^{2}\epsilon_{m}^{2}}{2}
 \left[\frac{1}{\epsilon_{m}\gamma_{m}}+\frac{1}{\epsilon_{d}\gamma_{d}}\right]\ .
\end{eqnarray*}

Taking Eq.~\ref{eq:Reflection_general_Prestep} from the derivation
of the complex reflection coefficient, one can further simplify\begin{eqnarray*}
\mathcal{A}_{m}\left(k_{SPP}R\right)\cdot\frac{\epsilon_{d}^{2}\epsilon_{m}^{2}}{2}\left[\frac{1}{\epsilon_{m}\gamma_{m}}+\frac{1}{\epsilon_{d}\gamma_{d}}\right] & = & \frac{\mathcal{B}_{m}\left(k_{SPP}R\right)}{2\pi\cdot\epsilon_{d}\cdot k_{SPP}}\cdot\int_{-\infty}^{\infty}\frac{H_{m}^{1}\left(\sqrt{\epsilon_{d}k_{0}^{2}-k_{z}^{2}}R\right)}{DH_{m}^{1}\left(\sqrt{\epsilon_{d}k_{0}^{2}-k_{z}^{2}}R\right)}\\
 &  & \sqrt{\epsilon_{d}k_{0}^{2}-k_{z}^{2}}\epsilon_{d}^{2}\epsilon_{m}^{2}\frac{\left(\gamma_{d}+\gamma_{m}\right)^{2}}{\left(k^{2}+\gamma_{d}^{2}\right)\left(k^{2}+\gamma_{m}^{2}\right)}dk_{z}\end{eqnarray*}
to
\begin{eqnarray*}
\mathcal{A}_{m}\left(k_{SPP}R\right) & = & \frac{\mathcal{B}_{m}\left(k_{SPP}R\right)}{\pi\cdot\epsilon_{d}\cdot k_{SPP}}\cdot\frac{\left(\gamma_{d}+\gamma_{m}\right)^{2}}{\frac{1}{\epsilon_{m}\gamma_{m}}+\frac{1}{\epsilon_{d}\gamma_{d}}}\int_{-\infty}^{\infty}\frac{H_{m}^{1}\left(\sqrt{\epsilon_{d}k_{0}^{2}-k_{z}^{2}}R\right)}{DH_{m}^{1}\left(\sqrt{\epsilon_{d}k_{0}^{2}-k_{z}^{2}}R\right)}\\
 &  & \frac{\sqrt{\epsilon_{d}k_{0}^{2}-k_{z}^{2}}}{\left(k^{2}+\gamma_{d}^{2}\right)\left(k^{2}+\gamma_{m}^{2}\right)}dk_{z}\ .\end{eqnarray*}

Using \begin{eqnarray*}
\frac{\left(\gamma_{m}+\gamma_{d}\right)^{2}}{\frac{1}{\epsilon_{m}\gamma_{m}}+\frac{1}{\epsilon_{d}\gamma_{d}}} & = & \sqrt{-\epsilon_{d}\epsilon_{m}}\cdot k_{SPP}^{3}\cdot\frac{\epsilon_{m}-\epsilon_{d}}{\epsilon_{m}+\epsilon_{d}}\ ,\end{eqnarray*}
which follows directly from the definitions of the $\gamma_{i}$,
one obtains\begin{eqnarray*}
\mathcal{A}_{m}\left(k_{SPP}R\right) & = & \frac{\mathcal{B}_{m}\left(k_{SPP}R\right)}{\epsilon_{d}\pi}\cdot\sqrt{-\epsilon_{d}\epsilon_{m}}\cdot k_{SPP}^{2}\cdot\frac{\epsilon_{m}-\epsilon_{d}}{\epsilon_{m}+\epsilon_{d}}\int_{-\infty}^{\infty}\frac{H_{m}^{1}\left(\sqrt{\epsilon_{d}k_{0}^{2}-k_{z}^{2}}R\right)}{DH_{m}^{1}\left(\sqrt{\epsilon_{d}k_{0}^{2}-k_{z}^{2}}R\right)}\\
 &  & \frac{\sqrt{\epsilon_{d}k_{0}^{2}-k_{z}^{2}}}{\left(k^{2}+\gamma_{d}^{2}\right)\left(k^{2}+\gamma_{m}^{2}\right)}dk_{z}\\
 & = & \frac{\mathcal{B}_{m}\left(k_{SPP}R\right)}{\epsilon_{d}\pi}\cdot\sqrt{-\epsilon_{d}\epsilon_{m}}\cdot\frac{\epsilon_{m}\epsilon_{d}}{\epsilon_{m}+\epsilon_{d}}\cdot\frac{\epsilon_{m}-\epsilon_{d}}{\epsilon_{m}+\epsilon_{d}}k_{0}^{2}\int_{-\infty}^{\infty}\frac{H_{m}^{1}\left(\sqrt{\epsilon_{d}k_{0}^{2}-k_{z}^{2}}R\right)}{DH_{m}^{1}\left(\sqrt{\epsilon_{d}k_{0}^{2}-k_{z}^{2}}R\right)}\\
 &  & \frac{\sqrt{\epsilon_{d}k_{0}^{2}-k_{z}^{2}}}{\left(k^{2}+\gamma_{d}^{2}\right)\left(k^{2}+\gamma_{m}^{2}\right)}dk_{z}\end{eqnarray*}
and further\begin{eqnarray*}
\pi\sqrt{-\epsilon_{d}\epsilon_{m}}\mathcal{A}_{m}\left(k_{SPP}R\right) & = & \mathcal{B}_{m}\left(k_{SPP}R\right)\cdot\epsilon_{p}^{2}\cdot\left(1-\frac{\epsilon_{m}}{\epsilon_{d}}\right)I_{m}^{\infty}\end{eqnarray*}
using the abbreviations \begin{eqnarray*}
I_{m}^{\infty} & = & \int_{-\infty}^{\infty}\frac{\sqrt{\epsilon_{d}-u^{2}}}{\left(u^{2}-\frac{\epsilon_{m}^{2}}{\epsilon_{m}+\epsilon_{d}}\right)\left(u^{2}-\frac{\epsilon_{d}^{2}}{\epsilon_{m}+\epsilon_{d}}\right)}\cdot\frac{H_{0}^{1}\left(\sqrt{\epsilon_{d}-u^{2}}\frac{\omega}{c}R\right)}{H_{1}^{1}\left(\sqrt{\epsilon_{d}-u^{2}}\frac{\omega}{c}R\right)}du\ ,\\
\epsilon_{p} & \equiv & \frac{\epsilon_{d}\epsilon_{m}}{\epsilon_{d}+\epsilon_{m}}\ .\end{eqnarray*}
Then, the reflection coefficient is given by\begin{eqnarray*}
r_{m} & = & \frac{\pi\sqrt{-\epsilon_{d}\epsilon_{m}}\cdot H_{m}^{1}\left(k_{SPP}R\right)-DH_{m}^{1}\left(k_{SPP}R\right)\epsilon_{p}^{2}\cdot\left(1-\frac{\epsilon_{m}}{\epsilon_{d}}\right)I_{m}^{\infty}}{-\pi\sqrt{-\epsilon_{d}\epsilon_{m}}\cdot H_{m}^{2}\left(k_{SPP}R\right)+DH_{m}^{2}\left(k_{SPP}R\right)\cdot\epsilon_{p}^{2}\cdot\left(1-\frac{\epsilon_{m}}{\epsilon_{d}}\right)I_{m}^{\infty}}\ .\end{eqnarray*}
This result has, as expected, similarities to the reflection of SPPs at a planar interface\cite{Gordon2006}.

%\end{widetext}
\twocolumngrid
%\bibliography{HankelReflection}
%Merlin.mbs v4.21 2009-07-09.
%

\end{document}